\documentclass[aps,prb,preprint,superscriptaddress,floatfix]{revtex4}

\usepackage{amsmath,amssymb,bm}
\usepackage{epsfig}
\usepackage{graphics}

\newcommand{\EP}{{\it e}{\rm -ph}}

\newcommand{\EE}{{\it e}{\it -e}}

\begin{document}

\title{Angle-resolved photoemission spectra of graphene from first-principles calculations}
\author{Cheol-Hwan Park}
\affiliation
{Department of Physics, 
University of California at Berkeley, 
Berkeley, California 94720, USA.}
\affiliation
{Materials Sciences Division, 
Lawrence Berkeley National Laboratory, Berkeley, 
California 94720, USA.}
\author{Feliciano Giustino}
\affiliation
{Department of Physics, 
University of California at Berkeley, 
Berkeley, California 94720, USA.}
\affiliation
{Materials Sciences Division, 
Lawrence Berkeley National Laboratory, Berkeley, 
California 94720, USA.}
\affiliation
{Department of Materials,
University of Oxford,
Parks Road, Oxford, OX1 3PH, UK.}
\author{Catalin D. Spataru}\affiliation
{Sandia National Laboratories,
Livermore, California 94551, USA.}
\author{Marvin L. Cohen} 
\affiliation
{Department of Physics, 
University of California at Berkeley, 
Berkeley, California 94720, USA.}
\affiliation
{Materials Sciences Division, 
Lawrence Berkeley National Laboratory, Berkeley, 
California 94720, USA.}
\author{Steven G. Louie}
\email{sglouie@berkeley.edu}
\affiliation
{Department of Physics, 
University of California at Berkeley, 
Berkeley, California 94720, USA.}
\affiliation
{Materials Sciences Division, 
Lawrence Berkeley National Laboratory, Berkeley, 
California 94720, USA.}

\date{\today}

\begin{abstract}
Angle-resolved photoemission spectroscopy (ARPES) is a powerful
experimental technique for directly probing electron dynamics in solids.
The energy vs.\ momentum dispersion relations and the associated spectral broadenings
measured by ARPES provide a wealth of information on quantum many-body interaction effects.
In particular, ARPES allows studies of the Coulomb interaction among electrons (electron-electron interactions)
and the interaction between electrons and lattice vibrations (electron-phonon interactions).
Here, we report {\it ab initio} simulations
of the ARPES spectra of graphene including both
electron-electron and electron-phonon interactions
on the same footing.
Our calculations reproduce some of the key experimental observations
related to many-body effects,
including the indication of a mismatch between the upper and lower halves of the Dirac cone.
\end{abstract}

\maketitle

In ARPES experiments a sample is illuminated
by monochromatic photons, which can extract electrons from the
sample if the photon energy exceeds the work function.
Analysis of the kinetic energy and angular distribution of the emitted electrons yields
the binding energy
of the electron in the material and its crystal momentum parallel to the
surface~\cite{damascelli:2003RMP_ARPES}.
The measured intensity $I({\bf k},\omega)$, where {\bf k}
and $\omega$ are the momentum and the binding energy of electrons
(usually referenced to the Fermi energy),
can be written as~\cite{damascelli:2003RMP_ARPES}
\begin{equation}
I({\bf k},\omega,\hat{e}_\nu,\hbar\nu)
=I_0({\bf k},\omega,\hat{e}_\nu,\hbar\nu)f(\omega) A({\bf k},\omega)\,,
\label{eq:I}
\end{equation}
where the function $I_0({\bf k},\omega,\hat{e}_\nu,\hbar\nu)$ takes into
account the absorption cross section of the incident photon of energy $\hbar\nu$
and polarization $\hat{e}_\nu$.
The function $f(\omega)$ is the Fermi-Dirac distribution, and $A({\bf k},\omega)$ is the
electron spectral function~\cite{damascelli:2003RMP_ARPES}. 
In interpreting ARPES measurements in a narrow energy range,
it is appropriate to assume that the absorption cross section
of the photon is constant. Under these conditions,
a measurement of the ARPES spectrum provides direct access to the electronic
spectral function $A({\bf k},\omega)$.
Within quantum many-body theory the electronic spectral function can be expressed as
\begin{equation}
A({\bf k},\omega)=\frac{2}{\pi}
\frac{-{\rm Im}\Sigma({\bf k},\omega)}
{\left[\omega-\varepsilon_{\bf k}-{\rm Re}\Sigma({\bf k},\omega)\right]^2
+\left[{\rm Im}\Sigma({\bf k},\omega)\right]^2
}\,,
\label{eq:A}
\end{equation}
where the
$\varepsilon_{\bf k}$'s are the the single-particle energy eigenvalues
of a reference mean-field system, and the self-energy $\Sigma({\bf k},\omega)$
accounts for the many-body interactions going beyond the mean-field
picture~\cite{damascelli:2003RMP_ARPES}.
For simplicity of discussion, the band indices are dropped from Eq.~(\ref{eq:A})
(see Methods).


In our investigation, we use density-functional Kohn-Sham eigenstates 
to describe the mean-field or non-interacting electrons.
The electron self-energy arising from the \EE\ interaction $\Sigma^{\EE}({\bf k},\omega)$ is evaluated within
the $G_0W_0$ approximation~\cite{hybertsen:1986PRB_GW} (Fig.~1a). This corresponds
to retaining the first diagram in the Feynman-Dyson perturbation expansion of the 
self-energy operator but in terms of the screened Coulomb interaction $W_0$
(Supplementary Discussion~3 for a comparison with previous
first-principles
calculations~\cite{trevisanutto:226405,attaccalite:arxiv}).
In this work, $G_0$ is constructed from the Kohn-Sham eigenvalues and eigenfunctions
of density-functional theory,
and $W_0$ is the bare Coulomb interaction screened by the full frequency-dependent
dielectric matrix $\epsilon({\bf r},{\bf r}',\omega)$ calculated within the random phase approximation.
The self-energy $\Sigma^{\EP}({\bf k},\omega)$ arising from the \EP\ interaction is similarly evaluated 
within the Migdal approximation~\cite{giustino:2007PRB_Migdal} (Fig.~1b).
With these choices, the \EE\ interaction and the \EP\ interaction
are described consistently within the same level of approximation~\cite{hedinlunqvist}
(see Methods).
The total self energy is then obtained as
\begin{equation}
\Sigma({\bf k},\omega)=\Sigma^{\EE}({\bf k},\omega)+\Sigma^{\EP}({\bf k},\omega)\,.
\label{eq:Sigma}
\end{equation}

Graphene,
a single layer of carbon atoms in a honeycomb structure,
has recently become an active research area in physics, 
chemistry, and nanoscience not only because
of its peculiar low-energy massless Dirac fermion band
structure~\cite{novoselov:2005Nat_Graphene_QHE,zhang:2005Nat_Graphene_QHE,berger:2006Sci_Graphene_Epitaxial},
but also because it holds promise for novel electronics and spintronics 
applications~\cite{wu:2007PRL_Graphene_mojunction}.
In particular, the epitaxial growth of graphene on silicon carbide (SiC)
has emerged as one of the promising routes towards large-scale production
of graphene~\cite{berger:2006Sci_Graphene_Epitaxial,wu:2007PRL_Graphene_mojunction}.

The interpretation of the measured ARPES spectra of epitaxial graphene
grown on silicon-rich surface of SiC has been controversial.
The spectral features observed in early ARPES measurements~\cite{bostwick:2007NatPhys}
were interpreted qualitatively in terms of \EE\ and \EP\ interactions.
On the other hand, experiments and analyses performed
by a different group suggested that the low-energy ARPES spectrum is dominated by a quasiparticle
energy gap of 0.2-0.3 eV at the Dirac point~\cite{zhou:2007NatMat}.
According to Ref.~\cite{zhou:2007NatMat}, this band gap likely
arises from the coupling of the graphene layer
with the reconstructed surface of the SiC substrate.
Despite a number of subsequent studies to resolve 
this controversy~\cite{rotenberg:2008NatMat_comment,zhou:2008NatMat_re}, the 
detailed nature of the low-energy quasiparticle dynamics in epitaxial graphene 
remains an open question.

The electronic structure and the photoemission spectra
of graphene have also been explored in a number of 
theoretical investigations using density-functional theory
approaches within a non-interacting single-particle
picture~\cite{mattausch:2007prl,varchon:126805,kim:176802}.
The effects of many-body interactions have also been investigated in
model calculations~\cite{polini:081411,hwang:081412}.
However, first-principles calculations of the full {\bf k}- and $\omega$-dependent
ARPES spectral function - which contains both the quasiparticle dispersions and their
lifetimes - including the \EE\ and the \EP\ interactions
have not been reported. The lack of first-principles many-body investigations
can partly be ascribed to the extremely demanding computational efforts
required to evaluate the real part of the electron self-energy, both for the \EE\
and for the \EP\ contributions.

In dealing with effects of \EE\ interactions, first-principles
calculations have advantages
over model calculations based on the two-dimensional
massless Dirac equation~\cite{polini:081411,hwang:081412}.
First, the scattering rate of even the low-energy charge
carriers, whose non-interacting dispersion relation can be well
approximated by the two-dimensional massless Dirac equation,
shows strong wavevector-anisotropic
behaviors~\cite{park:elel}.
This is because the carrier scattering rate
in graphene depends sensitively
on the sign of the band curvature~\cite{PhysRevB.59.R2474}.
Therefore, the significant wavevector anisotropy
in the electron scattering rate is not
captured by the model calculations.
Second, the model calculations require a cutoff of the
high-energy states and an adjustable parameter
mimicking effects of internal screening arising from the
high-energy states (including the $\pi$ states higher in energy
than the cutoff and the $\sigma$ states), in addition to
the external screening due to the environment.
The first-principles approach employing the full bandstructure
accounts for these processes explicitly and requires
neither a high-energy cutoff nor an empirical parameter
to describe the internal screening.


To determine the quasiparticle energy vs.\ momentum
dispersion relations from the calculated ARPES intensity maps
for graphene (Figs.~2a, 2d, and~2g), we follow the standard
procedure adopted in analyzing ARPES
experiments~\cite{damascelli:2003RMP_ARPES}.
First, the energy distribution curves (EDCs) are obtained
by performing cuts of the intensity maps at fixed photoelectron momentum 
(vertical cuts in Figs. 2a, 2d, and~2g).
Subsequently, the quasiparticle band structures are generated
by connecting the locations of the maxima in the EDCs for each
photoelectron momentum (Figs.~2b, 2e, and~2h).
This procedure ensures that the calculated and the measured
dispersion relations are obtained from the corresponding intensity maps
using the same procedure.


While the dispersion relations extracted from our
spectral functions for an isolated graphene layer are linear at large binding energy,
we observe a sizable kink for {\it n}-doped graphene near the Dirac point (${\bf k}=0$ in Fig.~2)
at an energy below the Dirac point energy.
Such kinks result in a mismatch between the linear extrapolations of 
the lower and the upper portions of the Dirac cone (Fig.~2). A similar phenomenon
has been observed in the measured ARPES spectra~\cite{bostwick:2007NatPhys,zhou:2007NatMat}.
In order to quantify the size of the kinks and the associated energy mismatch,
we have taken the energy difference between the two linear asymptotes of
the upper and the lower bands ($\Delta_{\rm kink}$ in Fig.~2f).
This calculated energy offset $\Delta_{\rm kink}$ is predominantly a result of many-body effects,
which comes from the {\it GW} self-energy, and is found to increase
with the doping level (see Fig.~3), consistent with the
experimental trend~\cite{bostwick:2007NatPhys}.
We have checked that, if many-electron effects are not considered,
the energy mismatch is several times smaller, with value $\le$~20~meV for the most heavily doped case
considered here and even smaller for other cases.
We note that, for the path along which the ARPES spectra are calculated (inset of Fig.~2c),
the nonlinearity of the bare graphene band is smallest.

Although the calculated $\Delta_{\rm kink}$ shows qualitative agreement with experiment,
after taking into account the screening of the SiC substrate
(Refs.~\cite{polini:081411,hwang:081412}, see Methods),
our calculated values
underestimate the experimentally observed offsets in Ref.~\cite{bostwick:2007NatPhys}
consistently by 60-90~meV; this discrepancy,
together with the comparison between theory and experiment of
electron linewidths discussed in Ref.~\cite{park:elel} and
later in this paper,
suggests the possible opening up of a band gap at the Dirac point energy
due to the interaction between graphene
and a reconstructed silicon-rich surface of
SiC~\cite{zhou:2007NatMat,mattausch:2007prl,varchon:126805,kim:176802}.
Our study shows the need of further first-principles studies,
considering both the atomistic structure of the graphene-substrate interface
and many-body effects.

The quasiparticle velocity can be extracted from the
simulated spectral functions. We find that
\EE\ interactions greatly enhance the band velocity by over 30~\%
compared to the density-functional theory value
in pristine graphene, but dielectric screening
from the SiC substrate (by weakening the \EE\ interaction)
reduces the quasiparticle velocity by as much as $\sim$10~\%
(Supplementary Discussion~1 and Fig.~\ref{SFig1}).
Moreover, the calculated velocity decreases with doping (Fig.~\ref{SFig1})
in agreement with previous calculations~\cite{attaccalite:arxiv}.


Our calculations also reveal phonon-induced kinks
near the Fermi energy at binding energies between
150 and 200 meV (e.\,g.\,, Fig.~4a) in good agreement with
experimental photoemission maps~\cite{bostwick:2007NatPhys,
zhou:2007NatMat,rotenberg:2008NatMat_comment,zhou:2008NatMat_re}.
These signatures of the \EP\ interaction in graphene
have been analyzed thoroughly both experimentally and
theoretically~\cite{park:2007PRL_Graphene_ElPh,park:2008PRB_Graphene_ElPh,calandra:205411,tse:236802}.

A complementary and important piece of information provided by ARPES intensity maps
is the linewidth of the electronic quasiparticle peaks. The linewidth $\Gamma_{n{\bf k}}$ is related to the lifetime
$\tau_{n{\bf k}}$ of the electron in a given quasiparticle state through $\tau_{n{\bf k}} = 2\hbar/\Gamma_{n{\bf k}}$,
and plays an important role in transport phenomena.
The electron linewidths are extracted from the measured photoemission spectra
by considering momentum distribution curves (MDCs), which are constant-energy cuts of
the intensity maps. The width of the MDC at a given binding energy can subsequently be obtained
using a Lorentzian fit~\cite{damascelli:2003RMP_ARPES}.
From our simulated ARPES spectra of graphene along the
$\Gamma$KM direction in wavevector space (Fig.~4a),
we obtained the MDCs (Fig.~4b)
and their widths (Fig.~4c).
Unlike previous methods~\cite{park:elel},  this procedure is direct and is not
confined to materials having a linear electronic dispersion.

The widths extracted from the calculated MDCs in {\it n}-doped graphene after including the effect of the
SiC substrate screening (see Methods) follow closely the
experimental measurements on epitaxial graphene
at large binding energies~\cite{bostwick:2007NatPhys}, although they underestimate
the experimental data close to the Dirac point energy (Fig.~4c). 
The agreement between calculated and measured widths
at large binding energy and the underestimation of the linewidths close
to the Dirac point energy provide additional support to the proposed scenario of a
band gap opening arising from the interaction with the SiC
substrate~\cite{zhou:2007NatMat,mattausch:2007prl,varchon:126805,kim:176802,park:elel}.
That is, the opening of a band gap
and the generation of midgap states near the Dirac point energy
(not included in our calculations)
would lead to increased linewidths in this energy regime~\cite{kim:176802}.
However, from the view point of theory, a conclusive statement can be made only after first-principles
calculations considering both the atomistic structure of the graphene-substrate
interface and the many-body effects, which is beyond the scope of this work.

The widths extracted from the calculated full ARPES spectrum and those obtained by
calculating the imaginary part of the on-shell electron self-energy
Im~$\Sigma({\bf k},\varepsilon_{\bf k})$~\cite{park:elel}
are similar for {\it n}-doped graphene with the same charge density.
(The charge density of {\it n}-doped graphene considered in Ref.~\cite{park:elel}
is different from that in Fig.~\ref{Fig4}.)
However, it is important to note that
the current method of using the full ARPES spectra,
although involving heavier computations, is more powerful
because it can be in general applied to systems whose bare electronic
energy dispersion is not linear.

In this work, the effects of substrate
optical phonons have not been considered.
These substrate phonons contribute to the
room-temperature transport properties of
graphene~\cite{graphene_limit}.
However, over a 1.5~eV range around the Fermi level
in doped graphene relevant to ARPES experiments~\cite{bostwick:2007NatPhys,zhou:2007NatMat},
the imaginary part of the electron self energy
that arises from interactions with the substrate
phonons is less than 1~meV~\cite{substrate_phonon},
about two orders
of magnitude smaller than the self-energy effects
intrinsic to graphene itself~\cite{park:elel}.

The present work shows that first-principles simulation
of ARPES spectra based on a quantum many-body theory approach
treating \EE\ and \EP\ interactions on the same footing
holds great potential for the interpretation of complex ARPES spectra.
In particular, a direct calculation of the quasiparticle spectral function
is needed to obtain meaningful comparisons with experimental data
(e.\,g.\,, the extraction of EDCs, MDCs, linewidths,
and quasiparticle dispersion relations).

\vspace{0.3in}
\noindent{\Large{\bf Methods}}
\vspace{0.2in}
\\
{\large{\bf 1. Computational setup}}
\\
The Kohn-Sham eigenstates of graphene
are obtained using density-functional theory calculations
within the local density approximation
(LDA)~\cite{ceperley:1980PRL_pseudopotential}
in a supercell geometry~\cite{PhysRevB.12.5575}.
Electronic wavefunctions in a $72\times72\times1$ k-grid
are expanded in a plane-waves basis
with a kinetic energy cutoff of 60~Ry.
The core-valence interaction is treated by means of {\it ab initio} norm-conserving
pseudopotentials~\cite{troullier:1991PRB_pseudopotential}.
Graphene layers are separated by 8.0~\AA\
and the Coulomb interaction is truncated to prevent
spurious interaction between periodic replicas~\cite{ismail-beigi:233103}.
We have checked that increasing the interlayer distance to
16.0~\AA\ makes virtually no difference in the calculated self energy.
Charge doping is modeled by an added electron density
with a neutralizing background.

Extending the procedure presented in Ref.~\cite{park:elel},
where only the imaginary part of the on-shell electron self-energy
Im$\Sigma^{\EE}(\varepsilon_{\bf k})$ arising from
electron-electron (\EE) interactions
is calculated
($\varepsilon_{\bf k}$ being the LDA energy eigenvalue),
we calculate the full frequency dependence of
both the real and the imaginary parts of the dielectric matrix
(within the random phase approximation) and
the self-energy operator $\Sigma^{\EE}(\omega)$ within the $G_0W_0$
approximation in the present work.
Thus our theory includes the two scattering mechanisms
arising from \EE\ interaction effects discussed in
previous model calculations performed
within the massless Dirac equation formalism~\cite{polini:081411,hwang:081412},
i.\,e.\,, electron-hole pair
and plasmon excitations.
Since our calculations are based
on first-principles, they are parameter free for suspended graphene
and can give information that depends on atomistic details,
e.\,g.\,, the effects of trigonal warping.
For convergence of the real part of $\Sigma^{\EE}(\omega)$,
we have included conduction bands with kinetic energy
up to 100 eV above the Fermi level.
The frequency dependent dielectric matrix $\epsilon_{{\bf G},{\bf G}'}({\bf q},\omega)$
is calculated within the random phase approximation using the LDA wavefunctions
on a regular grid of $\omega$
with spacing $\Delta\omega=$0.125~eV~\cite{benedict:085116},
and the dielectric matrix at energies in between frequency grid points
is obtained by a linear interpolation.
In the calculation of the polarizability, for numerical convergence,
an imaginary component of magnitude $\Delta\omega$ of 0.125~eV
is introduced in the energy denominator.
Convergence tests showed that the dimension
of the dielectric matrix may be truncated at
a kinetic energy cutoff of $\hbar^2G^2/2m=$12~Ry.
Additionally, we obtain the electron self-energy arising from
electron-phonon (\EP) interactions $\Sigma^{\EP}(\omega)$
following Ref.~\cite{park:2008PRB_Graphene_ElPh} for different levels of doping.

In our approach, we describe the \EE\ and the \EP\ interactions
within the same level of approximation. Previous studies
suggested the use of the quasiparticle dispersions renormalized 
by the \EE\ interaction to compute the \EP\
interaction~\cite{lazzeri:081406}. This latter procedure (not adopted here)
would correspond to including some of the higher order processes 
in $\Sigma^{\EP}(\omega)$ whilst neglecting them
in $\Sigma^{\EE}(\omega)$, and would result in an
unbalanced evaluation of \EE\ and \EP\ effects according
to different levels of approximation.
In any event, even if some \EP\ matrix elements
calculated after renormalizing the bands through
the {{\it GW}} approximation were 20$\%$ to 40$\%$ larger 
than those used in this work~\cite{lazzeri:081406},
the band velocity would also be enhanced by a similar factor
(this work and Refs.~\cite{trevisanutto:226405,attaccalite:arxiv}),
and the two factors would cancel out approximately
in the calculation of $\Sigma^{\EP}(\omega)$ 
[cf.\ Eq.~(2) of Ref.~\cite{park:2007PRL_Graphene_ElPh}].
Therefore, we estimate that the effects of such alterations
on our results are not significant.
\vspace{0.2in}
\\
{\large{\bf 2. Angle-resolved photoemission spectra}}
\\
First, the trace of the spectral function with respect to band index {\it n}, i.e.,
$A({\bf k},\omega) = c \sum_n A_{nn}({\bf k},\omega)$, is calculated. Here,
$c$ is a normalization constant, $\left|n\right>$ are the Kohn-Sham eigenstates, and
\begin{equation}
A_{nn}({\bf k},\omega) = \frac{2}{\pi}
\frac{-{\rm Im}\left<n{\bf k}|\Sigma(\omega)|n{\bf k}\right>}
{\left[\omega-\varepsilon_{\bf k}
-{\rm Re}\left<n{\bf k}|\Sigma(\omega)|n{\bf k}\right>\right]^2
+\left[{\rm Im}\left<n{\bf k}|\Sigma(\omega)|n{\bf k}\right>\right]^2}\,.
\label{eq:Ann}
\end{equation}

To simulate the measured angle-resolved photoemission spectra from the calculated
spectral functions, we multiplied the spectral function by the Fermi-Dirac distribution
$f(\omega)$ [Eq.~(1) of the manuscript] with $T=25$~K at which the
experiments were performed~\cite{bostwick:2007NatPhys,zhou:2007NatMat}.
Then, to take into account the experimental resolutions in energy and momentum,
we have convoluted the intensity maps with a two-dimensional
Lorentzian mask with $\Delta k$=0.01~\AA$^{-1}$ and $\Delta\omega=25$~meV,
corresponding to the experimental resolution~\cite{bostwick:2007NatPhys,zhou:2007NatMat}.
(This convolution results in finite linewidths even for zero binding-energy states.)
In simulating the photoemission spectra along the $\Gamma$KM direction
(Fig.~4a of the manuscript),
we have used only one branch of the two linear bands in order to simulate
the matrix element effects in
$I_0({\bf k},\omega,\hat{e}_\nu,E_\nu)$ [Eq.~(1) of the manuscript] 
(cf.\ Fig.~2 of Ref.~\cite{bostwick:2007NatPhys}).
In calculating the width of momentum distribution curves
(the linewidths in Fig.~4c),
we have arbitrarily subtracted off a constant from the simulated widths
so that the width vanishes at zero binding energy, as also done in
the analysis of experimental data in Ref.~\cite{bostwick:2007NatPhys}.
\vspace{0.2in}
\\
{\large{\bf 3. Substrate screening}}
\\
To include the effects of the dielectric screening associated with the
silicon carbide (SiC) substrate,
we have, as done in previous studies~\cite{bostwick:2007NatPhys,polini:081411,hwang:081412} (Supplementary Discussion~2 and Fig.~\ref{SFig2}),
renormalized the bare Coulomb interaction by an effective background dielectric constant 
$\epsilon_{\rm b}=(1+\epsilon_{\rm SiC})/2$, where $\epsilon_{\rm SiC}$=6.6
is the optical dielectric constant of silicon carbide~\cite{park:elel};
one takes the average of the vacuum dielectric constant
and the substrate dielectric constant because graphene is sandwiched in between
the two media~\cite{jackson}.
Along the lines of Ref.~\cite{bostwick:2007NatPhys},
we do not take into account atomistic interactions between
graphene and the reconstructed surface of the
silicon carbide substrate~\cite{mattausch:2007prl,varchon:126805,kim:176802}.

\vspace{0.3in}
\noindent{\Large{\bf Supplementary information}}
\vspace{0.2in}
\\
{\large{\bf 1. Velocity renormalization}}
\\
Our simulated spectral functions allow us to study the velocity of
Dirac fermions in the linear regime (away from the Dirac point)
as a function of dielectric screening of the substrate and doping (Fig.~\ref{SFig1}).
The band velocity of our model epitaxial graphene [$\epsilon_{\rm b}=(1+\epsilon_{\rm SiC})/2=3.8$]
is found to be smaller than that of
suspended graphene ($\epsilon_{\rm b}=1$) by as much as $\sim$10\% 
(Fig.~\ref{SFig1}).
In addition, the velocity decreases as doping increases (Fig.~\ref{SFig1}),
in agreement with the previous calculation~\cite{attaccalite:arxiv}.
Both trends are easily explained by observing that 
the polarizability of the substrate and the additional
electrons in the graphene layer both lead to weaker \EE\ interactions.
\vspace{0.2in}
\\
{\large{\bf 2. Substrate dielectric function}}
\\
In this section, we show that the dielectric
function of silicon carbide (SiC) $\varepsilon^{\rm SiC}({\bf q},\omega)$ is
well represented by the value at ${\bf q}=0$ and $\omega=0$
(i.\,e.\,, the optical dielectric constant) as far as our
calculation is concerned.
We also estimate that an error in the electron self energy
arising from this simplification is less than 10\%.

In calculating the imaginary part of the self-energy, the relevant
energy scale of the dielectric function of SiC is the
quasiparticle energy measured from the Fermi surface, since the
lifetime is determined by the real decay processes to lower energy
states [see, e.\,g.\,, Eq.~(5) of Ref.~\onlinecite{hwang:115434}].
The energy argument of the inverse dielectric function used in the
calculation of the imaginary part of the self energy at energy
$\omega$ varies between $\omega$ itself and the Fermi energy due
to the two Heaviside functions. As long as the dielectric function
is reasonably constant over this range, the results are valid.
Since we are only interested in the value of the linewidth for
states from the Fermi level down to about 2.5 eV below it
(as measured by experiment and presented in Figs.~2 and~4
of the main manuscript), we have checked the validity of our
approximation on substrate screening within this energy range
and the corresponding wavector range as discussed later.
The real part of the self energy is affected by
the dielectric screening involving large wavevectors; however,
the contribution coming from larger wavevector scatterings
is smaller because the Coulomb interaction decreases
with $q$. The error in the real part of the self energy arising
from the inaccuracy in the dielectric function for a larger
wavevector would be smaller than
that involving a smaller wavevector, which is estimated below.

In order to give a quantitative estimate of the possible error
arising from the frequency and wavevector dependence of the
dielectric function, we performed first-principles calculations
of the dielectric function of 3C-SiC showing dielectric responses
very similar to 6H-SiC~\cite{logothetidis:1996jap}, which is the
substrate used in
experiments~\cite{bostwick:2007NatPhys,zhou:2007NatMat}.
(The unit cell of 3C-SiC is much
smaller than that of 6H-SiC though.)


Supplementary Fig.~\ref{SFig2} shows that using the optical dielectric constant
($\omega=0$, ${\bf q}=0$) is a good approximation for the
energy and wavevector regime considered in our work:
the maximum variation in the dielectric function is 15\%.
(We have also checked that the inverse dielectric function
shows similar behaviors.)  Moreover, since (i) the self energy
is an average of many contributions, and, (ii) the finite frequency
and finite wavevector effects increase and decrease the value of
the dielectric function, respectively, the combined error coming
from our approximation will be less than 10\%.

Supplementary Fig.~\ref{SFig2} also shows (difference between solid and dashed lines)
that in fact the maximum anisotropy in the dielectric function of
SiC is $\sim$3\%.  (We have also checked that the
inverse dielectric function shows similar behaviors.) Since the
self energy is an average of all contributions, we expect that
the anisotropy in the calculated electron self energy arising
from that in the dielectric function of SiC will be
even smaller than that.
\vspace{0.2in}
\\
{\large{\bf 3. Comments on other calculations}}
\\
First-principles calculations
of the real part of the self-energy in graphene arising from
\EE\ interactions within the {\it GW} approximation
have been reported
previously~\cite{trevisanutto:226405,attaccalite:arxiv}.
The authors of Ref.~\onlinecite{attaccalite:arxiv} calculated the frequency 
dependence of the dielectric
matrices within the generalized plasmon-pole model. 
The authors of Ref.~\onlinecite{trevisanutto:226405} calculated
the full frequency dependence of the dielectric matrices
using the random phase approximation
as we did for the present work.
For consistency we compare our calculations to the latter study.

Our calculated velocity (1.23$\times10^6$~m/s) in suspended 
graphene is~$\sim$9\% larger than the one reported in Ref.~\onlinecite{trevisanutto:226405}
(1.12$\times10^6$~m/s).
Moreover, unlike the finding of Ref.~\onlinecite{trevisanutto:226405},
we do not observe a kink in the quasiparticle
band structure of suspended pristine graphene at an energy $\omega\sim-0.15$~eV
when \EP\ interactions are not included.
The fact that we observe
a gradual increase in the band velocity when approaching the Dirac point energy,
rather than a kink at a finite energy value is in line with
results of model analytical calculations~\cite{PhysRevB.59.R2474,sarma:121406}.
To clarify this difference, we observe that,
in the case of graphene, electronic states with wavevectors on nearest neighboring grid points 
of a discretized $N\times N$ mesh of the full Brillouin zone
have an energy difference $\Delta E~{\rm (eV)}\approx20/N$.
The Brillouin zone sampling adopted in Ref.~\onlinecite{trevisanutto:226405} ($10\times10$ points)
corresponds to electronic eigenstates with minimal energy separation of approximately 2 eV.
We find that this energy resolution is not sufficient to achieve convergence
in the dielectric matrices needed for the GW self energy (we used instead a $72\times72$ grid,
corresponding to energy separations of about 0.25~eV). Therefore we suggest that the difference
between the band dispersions and velocities calculated here and those of Ref.~\onlinecite{trevisanutto:226405}
may arise from the insufficient Brillouin zone sampling adopted in that work.

\vspace{0.3in}

\noindent{\bf Acknowledgments} The authors thank D.~M.~Basko,
E.~H.~Hwang, Y.-W. Son, A. Lanzara, and E.~Rotenberg
for fruitful discussions.
This work was supported by NSF Grant
No. DMR07-05941 and by the Director, Office of Science, Office of Basic Energy
Sciences, Division of Materials Sciences and Engineering Division,
U.S. Department of Energy under Contract No. DE- AC02-05CH11231.
Sandia is a multiprogram laboratory operated by Sandia Corporation,
a Lockheed Martin Company, for the US DOE.
Computational resources have been provided by TeraGrid and NERSC.

\newpage

  \begin{figure}
  \caption{
  {\bf Diagrams included in the calculated electron self-energy $\Sigma$.}
  {\bf a}, Electron self-energy $\Sigma^{\EE}=iG_0W_0$
  arising from \EE\ interactions within the $G_0W_0$ approximation.
  $G_0$ is the Green's function for bare electrons and $W_0$ is the screened Coulomb interaction.
  {\bf b}, Self-energy $\Sigma^{\EP}=ig^2G_0D$
  arising from \EP\ interactions within the Migdal approximation.
  Here, $g$ is the \EP\ interaction matrix element and $D$ is the dressed 
  phonon propagator.}
  \label{Fig1}
  \end{figure}

\newpage

  \begin{figure}
  \caption{{\bf Simulated ARPES spectra, energy distribution curves (EDCs) and quasiparticle band structures of suspended graphene including \EE\ and \EP\ interactions.}
  {\bf a}, Simulated ARPES spectrum of pristine graphene at $T=25$~K taken
  along the Brillouin zone segment indicated in the inset of {\bf c}.
  {\bf b}, EDCs extracted from {\bf a}. The central red curve corresponds to $k=0$ (the K point).
  {\bf c}, Quasiparticle band structure (solid red curve) obtained by connecting
  the peak positions of EDCs in {\bf b}.
  {\bf d} to {\bf f}, and {\bf g} to {\bf i}, Same quantities as in {\bf a} to {\bf c}
  for {\it n}-doped graphene with charge densities of $4.5\times10^{13}$~cm$^{-2}$
  and $1.2\times10^{14}$~cm$^{-2}$, respectively.
  The dashed blue lines in {\bf c}, {\bf f}, and {\bf i} indicate the asymptotes of the linear bands
  far from the Dirac point.
  The energy difference between the upper and the lower asymptotes close to the Dirac point
  is indicated by $\Delta_{\rm kink}$.}
  \label{Fig2}
  \end{figure}

\newpage

  \begin{figure}
  \caption{{\bf Mismatch between the upper and lower bands of the Dirac cone.}
  Calculated energy difference $\Delta_{\rm kink}$ between the asymptotic lines
  close to the Dirac point of the upper and the lower linear bands vs.\ doping
  for suspended graphene (red squares) and
  for graphene with a model dielectric screening (blue circles) corresponding
  to the SiC substrate (see Methods).
  The lines are a guide to the eye.}
  \label{Fig3}
  \end{figure}

\newpage

  \begin{figure}
  \caption{{\bf Momentum distribution curves (MDCs) of graphene and associated linewidths.}
  {\bf a}, Simulated ARPES spectrum of suspended {\it n}-doped graphene,
  for a doping level corresponding to a charge density of $4.5\times10^{13}$~cm$^{-2}$, taken
  along the Brillouin zone segment indicated in the inset of {\bf c}.
  {\bf b}, MDCs obtained from {\bf a}.
  {\bf c}, Width of the MDCs obtained from {\bf b} for suspended graphene (red curve) and
  that for graphene with a model dielectric screening corresponding to the SiC substrate
  (blue curve).
  The measured widths of the MDCs~\cite{bostwick:2007NatPhys} are shown
  for comparison (black curve).}
  \label{Fig4}
  \end{figure}

\newpage

  \begin{figure}
  \caption{{\bf Supplementary Fig. 1: Quasiparticle velocity in graphene.}
  The slopes (quasiparticle velocity) of the linear bands
  far from the Dirac point (indicated by the blue dashed lines in
  Figs.~1c, 1f, and~1i of the manuscript) vs.\ doping.
  Red squares and blue circles are calculated quantities for
  suspended graphene and for graphene with
  a model silicon carbide (SiC) dielectric screening, respectively.
  Black triangles are DFT results within the LDA.
  The lines are a guide to the eye.}
  \label{SFig1}
  \end{figure}

\newpage

  \begin{figure}
  \caption{{\bf Supplementary Fig. 2: Calculated dielectric functions of SiC versus energy.}
  Macroscopic dielectric function of 3C-SiC
  $\varepsilon^{\rm SiC}_{0,0}({\bf q},\omega)$ versus energy $\omega$.
  Quantities for different wavevectors are shown in different
  colors. Solid lines and dashed lines show quantities for the
  wavevector {\bf q} in two representative directions in wavevector
  space: $\Gamma$-X and $\Gamma$-L, respectively.}
  \label{SFig2}
  \end{figure}

\begin{thebibliography}{10}

\bibitem{damascelli:2003RMP_ARPES}
A. Damascelli, Z. Hussain, and Z.-X. Shen, Rev. Mod. Phys. {\bf 75},  473
  (2003).

\bibitem{hybertsen:1986PRB_GW}
M.~S. Hybertsen and S.~G. Louie, Phys. Rev. B {\bf 34},  5390  (1986).

\bibitem{trevisanutto:226405}
P.~E. Trevisanutto, C. Giorgetti, L. Reining, M. Ladisa, and V. Olevano, Phys.
  Rev. Lett. {\bf 101},  226405  (2008).

\bibitem{attaccalite:arxiv}
C. Attaccalite, A. Gr\"ueneis, T. Pichler, and A. Rubio, preprint available at
  http://arxiv.org/abs/0808.0786.

\bibitem{giustino:2007PRB_Migdal}
F. Giustino, M.~L. Cohen, and S.~G. Louie, Phys. Rev. B {\bf 76},  165108
  (2007).

\bibitem{hedinlunqvist}
L. Hedin and S. Lundqvist, {\em Solid State Physics} (Academic Press, New York,
  1969), Vol.~23, p.\ 1.

\bibitem{novoselov:2005Nat_Graphene_QHE}
K.~S. Novoselov, A.~K. Geim, S.~V. Morozov, D. Jiang, M.~I. Katsnelson, I.~V.
  Grigorieva, S.~V. Dubonos, and A.~A. Firsov, Nature {\bf 438},  197  (2005).

\bibitem{zhang:2005Nat_Graphene_QHE}
Y. Zhang, J.~W. Tan, H.~L. Stormer, and P. Kim, Nature {\bf 438},  201  (2005).

\bibitem{berger:2006Sci_Graphene_Epitaxial}
C. Berger, Z.~M. Song, X.~B. Li, X.~S. Wu, N. Brown, C. Naud, T.~B. Li, J.
  Hass, A.~N. Marchenkov, E.~H. Conrad, P.~N. First, and W.~A. {de Heer},
  Science {\bf 312},  1191  (2006).

\bibitem{wu:2007PRL_Graphene_mojunction}
X. Wu, M. Sprinkle, X. Li, F. Ming, C. Berger, and W.~A. {de Heer}, Phys. Rev.
  Lett. {\bf 101},  026801  (2008).

\bibitem{bostwick:2007NatPhys}
A. Bostwick, T. Ohta, T. Seyller, K. Horn, and E. Rotenberg, Nature Phys. {\bf
  3},  36  (2007).

\bibitem{zhou:2007NatMat}
S.~Y. Zhou, D.~A. Siegel, A.~V. Fedorov, and A. Lanzara, Nature Mater. {\bf 6},
   770  (2007).

\bibitem{rotenberg:2008NatMat_comment}
E. Rotenberg, A. Bostwick, T. Ohta, J.~L. McChesney, T. Seyller, and K. Horn,
  Nature Mater. {\bf 7},  258  (2008).

\bibitem{zhou:2008NatMat_re}
S.~Y. Zhou, D.~A. Siegel, A.~V. Fedorov, F.~E. Gabaly, A.~K. Schmid, A.~H.
  Castro{ }Neto, D.-H. Lee, and A. Lanzara, Nature Mater. {\bf 7},  259
  (2008).

\bibitem{mattausch:2007prl}
A. Mattausch and O. Pankratov, Phys. Rev. Lett. {\bf 99},  076802  (2007).

\bibitem{varchon:126805}
F. Varchon, R. Feng, J. Hass, X. Li, B.~N. Nguyen, C. Naud, P. Mallet, J.-Y.
  Veuillen, C. Berger, E.~H. Conrad, and L. Magaud, Phys. Rev. Lett. {\bf 99},
  126805  (2007).

\bibitem{kim:176802}
S. Kim, J. Ihm, H.~J. Choi, and Y.-W. Son, Phys. Rev. Lett. {\bf 100},  176802
  (2008).

\bibitem{polini:081411}
M. Polini, R. Asgari, G. Borghi, Y. Barlas, T. Pereg-Barnea, and A.~H.
  MacDonald, Phys. Rev. B {\bf 77},  081411  (2008).

\bibitem{hwang:081412}
E.~H. Hwang and S. Das{ }Sarma, Phys. Rev. B {\bf 77},  081412  (2008).

\bibitem{park:elel}
C.-H. Park, F. Giustino, C.~D. Spataru, M.~L. Cohen, and S.~G. Louie, Phys.
  Rev. Lett. {\bf 102},  076803  (2009).

\bibitem{PhysRevB.59.R2474}
J. Gonz\'alez, F. Guinea, and M.~A.~H. Vozmediano, Phys. Rev. B {\bf 59},
  R2474  (1999).

\bibitem{park:2007PRL_Graphene_ElPh}
C.-H. Park, F. Giustino, M.~L. Cohen, and S.~G. Louie, Phys. Rev. Lett. {\bf
  99},  086804  (2007).

\bibitem{park:2008PRB_Graphene_ElPh}
C.-H. Park, F. Giustino, J.~L. McChesney, A. Bostwick, T. Ohta, E. Rotenberg,
  M.~L. Cohen, and S.~G. Louie, Phys. Rev. B {\bf 77},  113410  (2008).

\bibitem{calandra:205411}
M. Calandra and F. Mauri, Phys. Rev. B {\bf 76},  205411  (2007).

\bibitem{tse:236802}
W.-K. Tse and S. Das{ }Sarma, Phys. Rev. Lett. {\bf 99},  236802  (2007).

\bibitem{graphene_limit}
J.-H. Chen, C. Jang, S. Xiao, M. Ishigami, and M.~S. Fuhrer, Nature Nanotech.
  {\bf 3},  206  (2008).

\bibitem{substrate_phonon}
S. Fratini and F. Guinea, Phys. Rev. B {\bf 77},  195415  (2008).

\bibitem{ceperley:1980PRL_pseudopotential}
D.~M. Ceperley and B.~J. Alder, Phys. Rev. Lett. {\bf 45},  566  (1980).

\bibitem{PhysRevB.12.5575}
M.~L. Cohen, M. Schl\"uter, J.~R. Chelikowsky, and S.~G. Louie, Phys. Rev. B
  {\bf 12},  5575  (1975).

\bibitem{troullier:1991PRB_pseudopotential}
N. Troullier and J.~L. Martins, Phys. Rev. B {\bf 43},  1993  (1991).

\bibitem{ismail-beigi:233103}
S. Ismail-Beigi, Phys. Rev. B {\bf 73},  233103  (2006).

\bibitem{benedict:085116}
L.~X. Benedict, C.~D. Spataru, and S.~G. Louie, Phys. Rev. B {\bf 66},  085116
  (2002).

\bibitem{lazzeri:081406}
M. Lazzeri, C. Attaccalite, L. Wirtz, and F. Mauri, Phys. Rev. B {\bf 78},
  081406(R)  (2008).

\bibitem{jackson}
J.~D. Jackson, {\em Classical Electrondynamics}, 3rd  ed. (Wiley, New York,
  1998).

\bibitem{hwang:115434}
E.~H. Hwang, B.~Y.-K. Hu, and S. Das{ }Sarma, Phys. Rev. B {\bf 76},  115434
  (2007).

\bibitem{logothetidis:1996jap}
S. Logothetidis and J. Petalas, J. App. Phys. {\bf 80},  1768  (1996).

\bibitem{sarma:121406}
S. Das{ }Sarma, E.~H. Hwang, and W.-K. Tse, Phys. Rev. B {\bf 75},  121406
  (2007).

\end{thebibliography}
\end{document}